# LYMAN-ALPHA SPECTROSCOPY IN NON-COMMUTATIVE SPACE-TIME

MUSTAFA MOUMNI

*Department of Matter Sciences, Biskra University, BP 145*
*Biskra, 07000, Algeria*
*m.moumni@univ-biskra.dz*

ACHOR BENSLAMA

*Physics Department, University of Constantine1*
*Constantine, 25000, Algeria*
*a.benslama@yahoo.fr*

SLIMANE ZAIM

*Department of Matter Sciences, Batna University*
*Batna, 05000, Algeria*
*slimane.zaiem@univ-batna.dz*

We study space-time non-commutativity applied to the hydrogen atom and the corrections induced to transitions frequencies. By writing the Dirac equation for noncommutative Coulomb potential, we compute noncommutative corrections of the energy levels using perturbation methods and by comparing to the Lamb shift accuracy we get a bound on the parameter of non-commutativity. We use this bound to study the effects on the Lyman-α ray and by induction look the possible influence of non-commutativity on some astrophysical and cosmological phenomena.

## 1. Introduction

Recently, there has been a certain amount of activity around the theme of cosmological and astrophysical applications of noncommutative geometry models of particle physics [1-7]. This work is in the same context and we are interested in the effects of space-time non-commutativity on the hydrogen spectroscopy and especially on the Lamb shift, the Lyman-α line.

The Lyman-α emission line is produced by recombination of electrons with ionized hydrogen atoms. The Lyman-α line and related diagnostics are now routinely used in numerous astrophysical studies for example to probe massive star formation across the Universe, to study properties of the interstellar and intergalactic medium across cosmic times, and to search for the most distant galaxies in the Universe [8].

In astronomy, a Lyman-Alpha Blob (LAB) is a huge concentration of a gas emitting the Lyman-α emission line [8]. LABs are some of the largest known individual objects in the Universe. Some of these gaseous structures are more than 400000 light years across. So far they have only been found in the high-redshift universe because of the ultraviolet nature of the Lyman-α emission line. Lyman-α Blobs may hold valuable clues for scientists to determine how galaxies are formed.

In astronomical spectroscopy, the Lyman-α forest is the sum of absorption lines arising from the Lyman-α transition of the neutral hydrogen in the spectra of distant galaxies and quasars [9]. These absorption lines result from inter-galactic gas through which the galaxy or quasar's light has travelled. The forest is created by the fact that photons that come to us from distant light sources show Hubble redshift that depends on the distance between us and the source of light. Since neutral hydrogen clouds at different positions between Earth and the distant light source see the photons at different wavelengths (due to the redshift), each individual cloud leaves its fingerprint as an absorption line at a different position in the spectrum as observed on Earth.

The Lyman-α forest is an important probe of the intergalactic medium and can be used to determine the frequency and density of clouds containing neutral hydrogen, as well as their temperature. Searching for lines from other elements like helium, carbon and silicon, the abundance of heavier elements in the clouds can also be studied.

The idea of taking space-time coordinates to be noncommutative goes back to the thirties of the last century. The goal was that the introduction of a noncommutative structure to space-time at small length scales could introduce an effective cut off which regularize divergences in quantum field theory. However this theory was plagued with several problems such as the violation of unitarity and causality, which make people abandon it. However noncommutative geometry was pursued on the mathematical side and especially with the work of Connes in the eighties of the last century [10].

In 1999, the interest for noncommutative geometry is renewed by the work of Seiberg and Witten on string theory [11]. They showed that the dynamics of the endpoints of an open string on a D-brane in the presence of a magnetic background field can be described by a Yang-Mills theory on a noncommutative space-time.

Noncommutative space-time is a deformation of the ordinary one in which the coordinates are promoted to Hermitian operators which do not commute:

$$\left[ x_{nc}^{\mu}, x_{nc}^{\nu} \right] = i\theta^{\mu\nu} = i\, C^{\mu\nu} \big/ (\Lambda_{nc})^2 \,;\, \mu,\nu = 0,1,2,3 \qquad (1)$$

where $\theta^{\mu\nu}$ is a deformation parameter and nc indices denote noncommutative coordinates. Ordinary space-time is obtained by making the limit $\theta^{\mu\nu} \to 0$. The noncommutative parameter is an anti-symmetric real matrix. $\Lambda_{nc}$ is the energy scale where the noncommutative effects of the space-time will be relevant and $C^{\mu\nu}$ are dimensionless parameters. A well-documented review can be found in [12-13].

We are interested in the noncommutative effects on the hydrogen atom spectroscopy. We start by writing the Dirac equation for the H-atom in framework of noncommutative space-time. Then, we compute the corrections of the energy levels induced by non-commutativity and get a limit on the noncommutative parameter from the 2P-2S Lamb shift (or the 28cm line) theoretical accuracy. We use this bound to estimate the contribution to the Lyman-α line.

## 2. Noncommutative Spectrum of Hydrogen Atom

We work on the space-time version of non-commutativity; thus instead of (1), we use:

$$\left[ x_{st}^j, x_{st}^0 \right] = i\theta^{j0} \tag{2}$$

the *st* subscripts are for noncommutative space-time coordinates. The 0 denotes time and *j* is for space coordinates.

As a solution to these relations, we choose the transformations:

$$x_{st}^j = x^j + i\theta^{j0}\partial_0 \tag{3}$$

Of course, the usual space coordinates $x^j$ satisfy the usual canonical permutation relations. For convenience, we use the vectorial notation as a compact writing:

$$\vec{r}_{st} = \vec{r} + i\vec{\theta}\partial_0 ; \vec{\theta} \equiv (\theta^{10}, \theta^{20}, \theta^{30}) \tag{4}$$

We have the Dirac equation (where $\alpha_i = \gamma_0 \gamma_i$ and the $\gamma_\mu$ are Dirac matrices):

$$i\hbar \partial_0 \psi = H\psi = \left[ (\vec{\alpha} \cdot \vec{p}) + m\gamma^0 - eA_0 \right]\psi \tag{5}$$

Because the kinetic energy does not change since it depends on the momentum $p$ which remains unchanged, we consider only changes on the potential energy by taking the Coulomb potential and constructing its noncommutative image. To achieve this, we write it with the new coordinates in the same way as the usual one and we restrict ourselves to the 1st order in $\theta$ because of the smallness of this parameter [12]:

$$A_0^{nc} = \frac{e}{r_{st}} = e\left[ \left( \vec{r} + i\vec{\theta}\partial_0 \right)^2 \right]^{-1/2} = \frac{e}{r}\left( 1 - i\frac{\vec{r} \cdot \vec{\theta}}{r^2}\partial_0 \right) + O(\theta^2) \tag{6}$$

As we work in spherical coordinates, an adequate choice of the noncommutative parameter is $\vec{\theta} = \theta^{0r}\vec{r}/r$; It is equivalent to that written in [6] in the case of noncommutative space-space and in [7] for the space-time case. This choice allows us to write the noncommutative Coulomb potential as (we note $\theta^{0r} = \theta_{st}$):

$$A_0^{nc} = \frac{e}{r}\left( 1 - i\frac{\theta_{st}}{r}\partial_0 \right) + O(\theta^2) = \frac{e}{r} - \frac{eE\theta_{st}}{\hbar}\frac{1}{r^2} + O(\theta^2) \tag{7}$$

We have use the fact that $i\hbar\partial_0\psi = E\psi$. The Hamiltonian can now be expressed as:

$$H = \left[ (\vec{\alpha} \cdot \vec{p}) + m\gamma^0 - e^2/r \right] + \left[ e^2 (E/\hbar) \theta_{st}/r^2 \right] = H^{(0)} + H^{(nc)} \tag{8}$$

$H^{(0)}$ is the Dirac Hamiltonian in the usual relativistic theory and $H^{(nc)}$ is the noncommutative correction to this Hamiltonian. The smallness of the parameter $\theta$ allows us to consider noncommutative corrections with perturbation methods; to the 1st order in $\theta$, the corrections of the eigenvalues are:

$$\Delta E^{(nc)} = \left\langle H^{(nc)} \right\rangle = e^2 E/\hbar \left\langle r^{-2} \right\rangle \theta_{st} \tag{9}$$

For calculating this average, simply use the expressions of Dirac solutions for hydrogen atom based on special functions; the work has already been done in [14] and the result is:

$$\left\langle \frac{1}{r^2} \right\rangle = \frac{2\kappa(2\kappa\varepsilon - 1)(1-\varepsilon^2)^{3/2}}{\alpha\sqrt{\kappa^2 - \alpha^2}\left[4(\kappa^2 - \alpha^2) - 1\right]} \left(\frac{mc}{\hbar}\right)^2 \left(\frac{1}{a_0^2}\right) \quad (10)$$

where $a_0 = \hbar^2/me^2$ is the 1st Bohr radius and $\varepsilon = E/mc^2$. Here $E$ represents the usual Dirac energy:

$$E_{n,j} = mc^2 \left\{ 1 + \alpha^2 \left[ (n-j-\tfrac{1}{2}) + \sqrt{(j+\tfrac{1}{2})^2 - \alpha^2} \right]^{-2} \right\}^{-1/2} \quad (11)$$

$\alpha = e^2/\hbar c$ is the fine structure constant and $j = l \pm 1/2$ is the quantum number associated to the total angular momentum. The number $\kappa$ is giving by the two relations: $\kappa = -j - 1/2$ if $j = l + 1/2$ and $\kappa = j + 1/2$ if $j = l - 1/2$. We see from (9) and (10) that through $\kappa$, the energy depends not only on the value of $j$ but also on the manner to get this value (or on $l$), unlike the usual Dirac energies in (11) which is the same for all the possible ways to obtain $j$; It implies that the non-commutativity remove the degeneracy $j = l + 1/2 = (l+1) - 1/2$ in hydrogen atom (states like $nP_{3/2}$ and $nD_{3/2}$).

We are interested in the study the Lyman lines, so we compute the corrections to the levels $n = 1; 2$:

$$\begin{aligned}
\Delta E^{(nc)}(1S_{1/2}) &= 1.065084 \cdot 10^{-4} \left(m^3 e^2 c^4 / \hbar^3\right) \theta \\
\Delta E^{(nc)}(2S_{1/2}) &= 1.331426 \cdot 10^{-5} \left(m^3 e^2 c^4 / \hbar^3\right) \theta \\
\Delta E^{(nc)}(2P_{1/2}) &= 0.443805 \cdot 10^{-5} \left(m^3 e^2 c^4 / \hbar^3\right) \theta \\
\Delta E^{(nc)}(2P_{3/2}) &= 0.443765 \cdot 10^{-5} \left(m^3 e^2 c^4 / \hbar^3\right) \theta
\end{aligned} \quad (12)$$

In order to make the dependence more visible, we do the development of the corrections from (9) and (10) to the 2nd order in $\alpha$; we find two expressions depending on the value of $\kappa$:

$$\begin{aligned}
\Delta E^{(nc)}_{n,j=l+1/2} &= \frac{m^3 e^2 c^4 \alpha^2}{jn^3 \hbar^3} \left[ 1 + \left( \frac{6j^2 + 6j + 1}{j(j+1)(2j+1)^2} + \frac{3}{(2j+1)n} - \frac{10j+9}{4(j+1)n^2} \right) \alpha^2 \right] \theta \\
\Delta E^{(nc)}_{n,j=l-1/2} &= \frac{m^3 e^2 c^4 \alpha^2}{(j+1)n^3 \hbar^3} \left[ 1 + \left( \frac{6j^2 + 6j + 1}{j(j+1)(2j+1)^2} + \frac{3}{(2j+1)n} - \frac{10j+1}{4jn^2} \right) \alpha^2 \right] \theta
\end{aligned} \quad (13)$$

As mentioned before, we see that space-time non-commutativity acts like the Lamb Shift and it removes the degeneracy $j = l + 1/2 = (l+1) - 1/2$ in usual hydrogen atom which characterizes the Coulomb potential.

From the precedent expressions, we get noncommutative correction to the Lamb Shift:

$$\Delta E^{(nc)}_{LambShift} = \Delta E^{(nc)}_{n,j=(l+1)-1/2} - \Delta E^{(nc)}_{n,j=l+1/2}$$
$$= \frac{m^3 e^2 c^4 \alpha^2}{j(j+1)n^3 \hbar^3} \left[ 1 + \left( \frac{6j^2 + 6j + 1}{j(j+1)(2j+1)^2} + \frac{3}{(2j+1)n} - \frac{2}{n^2} \right) \alpha^2 \right] \theta \quad (14)$$

We compute the correction for the $2P_{1/2} - 2S_{1/2}$ Lamb shift ($n = 2$ and $j = 1/2$) from (14) (or from the general relations in (11)) and we compare this result to the current theoretical accuracy for the Lamb shift $0,08\ kHz$ from [15]; we find the bound:

$$\Delta E^{(nc)}_{LambShift} \leq 0.08\,kHz \Rightarrow \theta_{st} \leq 3.254 \cdot 10^{-23}\,eV^{-2} \approx (0.18\,TeV)^{-2} \quad (15)$$

It is better than the one obtained in [16].

We take this limit as a reference and use it to study the noncommutative effects on the Lyman-α lines; we find:

$$\Delta E^{nc}(Ly\alpha_1) \approx \Delta E^{nc}(Ly\alpha_2) = 3.067 \times 10^{-12}\,eV$$
$$\Rightarrow \Delta \lambda^{nc}(Ly\alpha_1) \approx \Delta \lambda^{nc}(Ly\alpha_2) = 5.82 \times 10^{-11}\,Å \quad (16)$$

From the National Institute of Standards and Technology [17], the accuracy of the Ly-α wavelength is $1.18 \cdot 10^{-9}$Å, we see that N.C corrections are about 5% smaller; we conclude that, in the current status of experimental limits and taking as reference the bound of the noncommutative parameter coming from the Lamb shift, the effects of non-commutativity on Lyman-α line are negligible.

For the redshift and using its definition, we compute the noncommutative corrections as follows:

$$1 + z = \frac{\lambda_{obs}}{\lambda_{emit}} \Rightarrow \Delta^{(nc)} z = \frac{\Delta^{(nc)} \lambda}{\lambda_{emit}} \approx 5 \times 10^{-14} \quad (17)$$

And this can be translated into corrections on cosmological distance (again using its definition):

$$\Delta^{(nc)} d = \left( \Omega_M (1+z)^3 + \Omega_\Lambda (1+z)^2 + \Omega_k \right)^{-1/2} \frac{c}{H_0} \Delta^{(nc)} z \quad (18)$$

It is clear that the corrections are inversely proportional to the values of z. The maximum value for these corrections is achieved for $z \to 0$. If we take values of the parameters from the ΛCDM model as an example (for a review on ΛCDM, one can see [18]), we get for this limit:

$$\Delta^{(nc)} z \approx 5 \cdot 10^{-14} \Rightarrow \Delta^{(nc)} d \approx 5 \cdot 10^{-4}\,ly \quad (19)$$

And we can clearly see that noncommutative corrections are insignificant compared to both galactic and cosmological distances.

## 3. Conclusion

In this work, we look for space-time noncommutative hydrogen atom and induced phenomenological effects; for this we use the Bopp's shift formulation. We found that applying space-time non-commutativity to the electron in the H-atom modifies the Coulomb potential to give us the potential of Kratzer.

By solving the Dirac equation, we have calculated the corrections induced to energy levels. The space-time noncommutative corrections to the Dirac theory of hydrogen atom remove the degeneracy of the Dirac energies with respect to the total angular momentum quantum number $j = l + 1/2 = (l + 1) - 1/2$ in addition to the degeneracy of the Bohr energies with respect to the orbital quantum number $l$, and the energies are labeled $E_{n;j;l}$; non-commutativity acts like a Lamb shift here. This is explained by the fact that Lamb correction can be interpreted as a shift of $r$ in the Coulomb potential due to interactions of the bound electron with the fluctuating vacuum electric field, and non-commutativity is also a shift of $r$ as we can see from the Bopp's shift (3) and (4).

By comparing to theoretical limit of the Lamb shift, we get a bound on the parameter of non-commutativity. This bound is not affected if we use the two-particle Dirac equation because we have demonstrated in [19] that this introduces a minimal adjustment by a factor equal to $(1 + m_e/m_p)^{-3/2} = 0.99918$ to our computations of the corrections. So we use this limit to see the noncommutative effects on the Lyman-α line and found that they are about 5% smaller than the actual experimental accuracy; the noncommutative corrections on the Lyman-α line are negligible up to now.

Similarly, we calculated the corrections induced by non-commutativity on the redshift and, consequently, on astronomical distances. We found that they are all the smaller for large $z$ and that the maximum value is achieved for $z \to 0$: $\Delta z \approx 5 \times 10^{-14}$. And we conclude that noncommutative corrections are insignificant compared to both galactic and cosmological distances and they do not affect phenomena associated with this line.

**Acknowledgments**

This work was done as part of the PNR project (8/P/461) entitled "Theories of NonCommutative Fields from Matrix Models and Emerging Physics" under the patronage of the Ministry of Higher Education and Scientific Research of the Democratic and Popular Republic of Algeria.
Authors would like to thank the Laboratory of Mathematical and Subatomic Physics (LPMPS) of Constantine University for supporting this work.